\documentclass[aps,prl,twocolumn,groupedaddress,showpacs]{revtex4}

\usepackage{graphicx}
\usepackage{dcolumn}
\usepackage{bm}
\usepackage{amssymb}

\begin{document}

\title{Comment on ``Onset of Boson Mode at the Superconducting Critical
Point of Underdoped YBa$_2$Cu$_3$O$_y$"}

\author{Yoichi Ando}

\affiliation{Institute of Scientific and Industrial Research, 
Osaka University, Ibaraki, Osaka 567-0047, Japan}

\date{\today}

\pacs{74.25.Fy, 74.72.Bk}

\maketitle

In a recent Letter \cite{newPRL}, Doiron-Leyraud {\it et al.} reported a
study of low-$T$ thermal conductivity $\kappa$ of underdoped
YBa$_2$Cu$_3$O$_y$ (YBCO) across the superconductor-insulator (S-I)
boundary and argued for a novel bosonic mode as well as
charge-neutral fermions. In this Comment, we show that both of these
claims are dubious.

{\it Bosons.}---In Ref. \cite{newPRL}, a larger $\kappa$ in insulator
compared to that in superconductor was interpreted to be due to a novel
bosonic mode. However, it is important to notice that the
crystallographic structure of YBCO changes from orthorhombic to
tetragonal at this S-I boundary \cite{Rosat-Mignot,note}. Since the
crystals studied in Ref. \cite{newPRL} were not detwinned, there were
twin boundaries, which scatter phonons, in the orthorhombic
superconducting samples; also, the ``tetragonal" YBCO near the S-I
boundary is known to be locally orthorhombic \cite{Jorgensen}, and the
associated orthorhombic strains would cause growing phonon scatterings
towards the S-I boundary. Therefore, it is likely that the decrease in
$\kappa$ upon approaching the S-I boundary in the insulating regime was
a reflection of the phonon thermal conductivity $\kappa_p$ being
affected by the growth of orthorhombic strains until twin boundaries are
established; indeed, in Fig. 1 of the Letter, the $\kappa/T$ data for
4.7\% and 5.4\% are only $\sim$1/4 of that for $y$ = 6.0 at 0.3 K,
suggesting strong phonon scatterings. In this regard, it should be
noted that in relation to the discrepancy between the data of Refs.
\cite{newPRL,Sutherland} and those of Ref. \cite{Sun1}, the Letter
misinformed the readers that the samples used in the latter were dirtier
because of the growth in ``zirconia" crucibles and this could be the
source of the discrepancy. In reality, the samples of Ref. \cite{Sun1}
were grown in pure Y$_2$O$_3$ crucibles, and those crystals have been
documented \cite{Sun2,Sun3} to be as clean as those grown in BaZrO$_3$
crucibles. The difference that is relevant to the present issue is that
all the samples used in Ref. \cite{Sun1} were detwinned \cite{note} and
were free from the complications described here.

{\it Fermions.}---The Letter analyzed the data by postulating that the
fermionic contribution $\kappa_f$ simply behaves as $aT$ and concluded
the existence of charge-neutral fermions. However, one must remember
that $\kappa_f = aT$ is valid only when the scattering rate $\Gamma$ of
fermions is constant. Irrespective of whether charge-neutral fermions
exist, certainly there are charge-carrying electrons whose $\Gamma$
keeps changing (because of the $\log(1/T)$ resistivity divergence),
which means that the limit $\kappa_f = aT$ is not achieved \cite{Sun1}.
One might think that the contribution of the charge carriers,
$\kappa_e/T$, would be negligible in ``insulating" samples, but in the
present case $\kappa_e/T$ is actually expected to be sizable: an
estimate for the 4.7\% sample using the Wiedemann-Franz law (WFL)
suggests $\kappa_e/T$ of the order of 20 $\mu$WK$^{-2}$cm$^{-1}$ at 90
mK, which is $\sim$20\% of total $\kappa/T$. Note that the WFL would
only give an order-of-magnitude estimate here, because it must be
violated in the localization regime \cite{Sun1}; however, in cuprates
the violation of WFL tends to give bigger $\kappa_e/T$ than is expected
from WFL \cite{Proust}, only to magnify the importance of $\kappa_e/T$
\cite{note1}. In any case, given that one cannot assume $\kappa_f = aT$
and that even $\kappa_p$ is likely to be changing upon time doping due
to the structural changes, it is essentially impossible to separate
$\kappa_f$ from $\kappa_p$ in the present case; in this regard, taking
the difference between two doping states of the same sample is of little
use when $\kappa_p$ changes upon time doping. Actually, if the analysis
in the Letter regarding the fermions were correct, it implies that
charge-carrying electrons in the normal state are {\it completely
incapable of carrying heat} even when they have a reasonably large
charge conductivity ($\sim 10^3$ $\Omega^{-1}$cm$^{-1}$), which is a
{\it very} strange situation.

In conclusion, the existence of a novel bosonic mode is questionable
because the Letter fails to address the effects of structural changes
near the S-I boundary. Moreover, the conclusion regarding charge-neutral
fermions is doubtful because the analysis is not valid.

We thank Kouji Segawa and X. F. Sun for useful discussions. 
This work was supported by KAKENHI 16340112 and 19674002.


\begin{thebibliography}{}

\bibitem{newPRL}
N. Doiron-Leyraud {\it et al.}, Phys. Rev. Lett. {\bf 97}, 207001
(2006).

\bibitem{Rosat-Mignot}
J. Rosat-Mignot {\it et al.}, Physica B (Amsterdam) {\bf 169}, 58
(1991).

\bibitem{note}
As prepared, insulating YBCO is tetragonal; however, application of
uniaxial pressure makes it weakly orthorhombic, which stays after
removing the pressure. The insulating samples studied in Refs.
\cite{Sun1,Ando} were in this uniaxial-pressure-induced orthorhombic
state.

\bibitem{Jorgensen}
J. D. Jorgensen {\it et al.}, Phys. Rev. B {\bf 41}, 1863 (1990).

\bibitem{Sutherland}
M. Sutherland {\it et al.}, Phys. Rev. Lett. {\bf 94}, 147004
(2005).

\bibitem{Sun1}
X. F. Sun, K. Segawa, and Y. Ando, Phys. Rev. B {\bf 72},
100502(R) (2005).

\bibitem{Sun2}
X. F. Sun, K. Segawa, and Y. Ando, Phys. Rev. Lett. {\bf 93} 
107001 (2004).

\bibitem{Sun3}
X. F. Sun {\it et al.}, Phys. Rev. Lett. {\bf 96}, 017008 (2006).

\bibitem{Proust}
C. Proust {\it et al.}, Phys. Rev. B {\bf 72}, 214511 (2005).

\bibitem{note1}
The discussion of WFL in the Letter is pointless because $\rho$
and $\kappa_0/T$ were asserted to be due to different carriers.

\bibitem{Ando}
Y. Ando {\it et al.}, Phys. Rev. Lett. {\bf 88}, 137005 (2002).


\end{thebibliography}
\end{document}